\documentclass[10pt,journal]{IEEEtran}

\ifCLASSOPTIONcompsoc

\usepackage[nocompress]{cite}
\else

\usepackage{cite}
\fi

\ifCLASSINFOpdf

\else

\fi

\usepackage{subfigure}
\usepackage{graphicx}
\usepackage{balance}
\usepackage{amsmath,amsthm, amssymb}
\usepackage{amssymb,wasysym}
\usepackage{amsfonts,balance}
\usepackage{mathrsfs}
\usepackage{cite}
\usepackage{color}
\usepackage{url}
\usepackage{bm}
\usepackage{wrapfig}
\usepackage{verbatim}
\usepackage{dsfont}
\usepackage{amsmath}
\usepackage{algorithm}
\usepackage{algpseudocode}
\usepackage{xurl}
 \usepackage[dvipsnames]{xcolor}
\usepackage{tabularx}
\usepackage{enumitem}
\usepackage{float}

\newcolumntype{C}[1]{>{\centering\arraybackslash}p{#1}}

\usepackage[colorlinks=true, linkcolor=black, citecolor=black, urlcolor=black, bookmarks=false]{hyperref}

\theoremstyle{definition}

\allowdisplaybreaks[3]

\ifodd 1
\else

\fi

\graphicspath{{figures/}}

\begin{document}

	\title{SpaceMoE: Towards Orbital General Intelligence with Distributed Mixture-of-Experts Inference}
	
	\author{Qian Chen,~\IEEEmembership{Member,~IEEE}, Xianhao Chen,~\IEEEmembership{Member,~IEEE}, \\ Min Sheng,~\IEEEmembership{Fellow,~IEEE}, and Kaibin Huang,~\IEEEmembership{Fellow,~IEEE}
		\thanks{Q. Chen, X. Chen, and K. Huang are with the Department of Electrical and
Computer Engineering, The University of Hong Kong, Hong Kong (Email: qchen@eee.hku.hk, xcheneee@hku.hk, and huangkb@hku.hk). M. Sheng is with the State Key Laboratory of Integrated Service Networks,
Institute of Information Science, Xidian University (XDU), Xi'an, Shaanxi,
China (Email: msheng@mail.xidian.edu.cn).
Corresponding authors: X. Chen.}
	}

	\markboth{}%
	{Shell \MakeLowercase{\textit{et al.}}: Bare Advanced Demo of IEEEtran.cls for IEEE Computer Society Journals}

	
		\IEEEtitleabstractindextext{
				\begin{abstract}
	 As satellite networks evolve to support increasingly diverse services and artificial general intelligence (AGI), large language models (LLMs) are emerging as a critical foundation for future space systems. However, deploying LLMs on satellites is hindered by stringent constraints on onboard memory, computation, and energy. In this context, the mixture-of-experts (MoE) architecture emerges as a promising solution, leveraging sparse expert activation to enable scalable model inference. By harnessing the architectural advantages of MoE, this article provides a comprehensive overview of SpaceMoE, a new paradigm for distributed MoE inference in satellite networks. We first review recent industrial progress and emerging standardization trends that motivate the evolution toward space AGI systems. Then, we introduce the fundamentals and architectural evolution of SpaceMoE. Subsequently, we discuss three fundamental design problems in SpaceMoE, namely expert placement, expert selection, and hidden-state transmission and routing, highlighting how satellite-specific factors such as dynamic topology, battery degradation, and thermal limits fundamentally reshape their solutions. Finally, we outline promising research directions for realizing scalable, efficient, and sustainable on-orbit MoE inference in future satellite networks.

					\end{abstract}

				}
	
	\maketitle

	\IEEEdisplaynontitleabstractindextext
	\IEEEpeerreviewmaketitle

\section{Introduction}

The IMT-2030 framework identifies ubiquitous connectivity as a core usage scenario for sixth-generation (6G) mobile networks, where satellite networks extend communication coverage to bridge the digital divide and connect the unconnected worldwide~\cite{8766143}. As they evolve into basic infrastructure, satellite networks are rapidly transitioning from communication infrastructure to integrated communication-AI infrastructure, ultimately enabling space general intelligence~\cite{10121575,11104147}. This shift is driven by the advent of large language models (LLMs)~\cite{10835069} and the growing need for intelligent operations across heterogeneous tasks and modalities, whether in space or on the ground. Specifically, future space systems must handle two primary types of LLM-powered services: \textit{terrestrial LLM services} for ground subscribers in infrastructure-poor areas, such as analyzing multimodal live video and human instructions to supervise port facilities; and \textit{on-orbit LLM services}, such as real-time reasoning and synthesis over terabits of raw remote sensing data to track dynamic environmental events~\cite{10045716}. By executing these advanced tasks directly in orbit, such as low-earth orbit (LEO) satellites, satellite networks can eliminate the significant latency and backhaul bottlenecks in traditional ground-centric AI architectures via satellite-to-ground backbone~\cite{chen2024space}.

While the vision of space general intelligence is promising, directly deploying LLMs on satellites remains highly challenging. According to the scaling law~\cite{kaplan2020scaling}, achieving superior performance in LLMs generally requires a continuous expansion of model size. This trend leads to substantial requirements for memory, computation, and energy resources, which are difficult to support on individual satellites with limited onboard hardware. 
 To address this challenge, the mixture-of-experts (MoE) architecture provides a promising alternative~\cite{jiang2024mixtral}. By leveraging sparse activation, MoE activates only a small subset of experts for each input token (e.g., Top-$K$ gating with $K$ experts activated per layer), thereby significantly reducing the per-token computation cost. Importantly, under a similar activated parameter budget, MoE often achieves superior model capacity and performance compared with dense counterparts, making it particularly attractive for resource-constrained satellite systems~\cite{11022729}. 
 By decoupling model capacity from computation cost, MoE provides a practical path toward realizing on-orbit intelligence under the stringent constraints of orbital platforms.

To unleash the full potential of MoE architecture in space, this article advocates \textit{SpaceMoE} as a new paradigm for distributed LLM inference in satellite networks. Although MoE leverages sparse activation to reduce computation, it does not alleviate the massive memory footprint\footnote{For example, Mixtral-8$\times$7B requires about 96 GB of memory, exceeding even high-end GPUs such as the A100 80 GB, while 45.1 B out of its 46.7 B parameters are experts.}. Overcoming this bottleneck necessitates distributed inference across multiple satellites. Consequently, SpaceMoE should be defined by two essential characteristics to make on-orbit operations feasible. \textit{First}, it should rely on a distributed MoE architecture, where experts are distributed across multiple satellites and dynamically selected during inference. \textit{Second}, SpaceMoE must enable orbit-aware resource management that adapts MoE inference to time-varying network topology and satellite-specific system constraints. Unlike terrestrial distributed MoE systems, SpaceMoE must operate under tightly coupled topological, energy, and thermal constraints that fundamentally reshape both the architecture and operation of distributed MoE inference. These unique challenges fundamentally reshape how distributed inference is executed, creating new challenges and opportunities.

To explore this emerging paradigm, this article first reviews the industrial and standardization trends driving on-orbit general intelligence toward distributed and scalable MoE inference in Section \ref{sec:industry}. Then, we present the fundamentals and architectural evolution of SpaceMoE in Section \ref{sec:architecture}. Building on this foundation, we identify three core design problems of SpaceMoE, discuss how each of them should be rethought in satellite networks, and highlight the distinctive tradeoffs in Section \ref{sec:prob_solution}. Finally, we outline the open research directions for the practical realization of SpaceMoE systems in Section \ref{sec:openprob} and conclude this article in Section \ref{sec:conclusion}.

\section{Industry Progress and Standardization for SpaceMoE}\label{sec:industry}
In this section, we review recent progress in deploying LLMs in satellite networks from two perspectives: industry progress and standardization trends. These developments collectively highlight the growing need for efficient and scalable LLM architectures in space, thereby motivating the discussion of SpaceMoE in the next section.

\subsection{Industrial Progress}

\subsubsection{Space AI Infrastructure}

A prerequisite for space general intelligence is the deployment of server-class hardware in orbital environments. In November 2025, Starcloud launched Starcloud-1, carrying an NVIDIA H100 GPU—an early but significant attempt to deploy high-performance AI accelerators in orbit. A more expansive infrastructure vision is embodied in Google's Project Suncatcher, which proposes a comprehensive space-based computing architecture integrating satellite constellations, TPU-based accelerators, laser inter-satellite links (ISLs), and solar-powered energy systems, with prototype satellite launches anticipated around 2027.

Concurrently, the continued rapid expansion of large-scale LEO constellations is creating new opportunities for distributed space computing. Building upon its established satellite infrastructure and laser ISL capabilities, recent regulatory filings by SpaceX suggest a prospective extension toward orbital data-center functionality. These proposals envision very large-scale satellite systems capable of supporting compute-intensive AI workloads, with solar energy identified as a key enabling resource.

\subsubsection{Space LLM Deployment}

In parallel with hardware infrastructure development, substantial progress has been made in deploying and validating foundation models in space. In September 2024, Guoxing Aerospace reported the first on-orbit validation of LLM inference. In May 2025, they further deployed a 12-satellite computing constellation, extending the scope from single-satellite validation to multi-satellite collaborative inference. Application-oriented validation is also underway. In September 2025, a traffic analysis model had been deployed by Guoxing Aerospace for remote sensing data processing. By November 2025, the Qwen3 foundation model had been deployed in orbit. Starcloud-1 reportedly supported inference workloads using Google's Gemma model and a variant of Gemini, with subsequent experiments extending to on-orbit training of nanoGPT.

\subsection{Standardization and Early Initiatives}

6G standardization is increasingly moving beyond connectivity-centric design toward an intelligence-oriented vision. In particular, the IMT-2030 framework identifies ubiquitous intelligence as one of the primary design principles for future mobile systems.

Within this broader intelligence-oriented evolution toward 6G, recent 3GPP standardization activities provide two important signals for SpaceMoE: the continuous enhancement of non-terrestrial network (NTN) capabilities and the growing integration of AI into communication systems. On the satellite side, Release 17 introduced the first normative NTN support, while Release 18 further enhanced New Radio (NR)-NTN capabilities for LEO satellite access and radio protocol operation. Release 19 continues this evolution toward more capable NTN architectures, including regenerative payloads with stronger onboard processing capabilities. These releases indicate that satellites are gradually evolving from bent-pipe relays into more integrated network nodes. In parallel, 3GPP has also advanced AI-integrated communication systems. Release 18 initiated AI support for next-generation radio access network (NG-RAN) and the NR air interface, covering use cases such as network energy saving, load balancing, and mobility optimization, while Release 19 further extends AI support in RAN. Together, these NTN and AI standardization efforts suggest that future satellite networks will not only provide ubiquitous connectivity, but also support intelligent onboard processing and distributed AI services.

Beyond formal standards, early industry initiatives also point to growing recognition of satellite-related AI scenarios. For example, GSMA’s Open Telco AI initiative already lists satellite as one of the available benchmark and tooling scenarios. Although this does not constitute a dedicated standard for satellite-based LLM deployment, it indicates that satellites are starting to enter telecom AI development frameworks.

\subsection{Takeaways}
Collectively, these industrial and standardization developments signal an evolution toward deploying AI models, including LLMs, in orbit. However, given the computation and energy constraints of satellites, deploying dense models are still very challenging. The MoE architecture is therefore uniquely suited to address this dilemma. In what follows, we introduce SpaceMoE as a feasible pathway to support space intelligence.

\section{Fundamentals and Architectural Evolution of SpaceMoE}\label{sec:architecture}

This section provides an overview of MoE inference in space environments from a system perspective. We first describe the basic procedure of MoE inference to illustrate the overall processing flow. Building on this, we then discuss the evolution from traditional space inference to SpaceMoE. Finally, we provide a comparative discussion that examines these solutions in resource-constrained satellite networks.

\subsection{Basic Procedure of MoE Inference}
\begin{figure}[t]
	\centering
    \includegraphics[width = 0.45\textwidth]{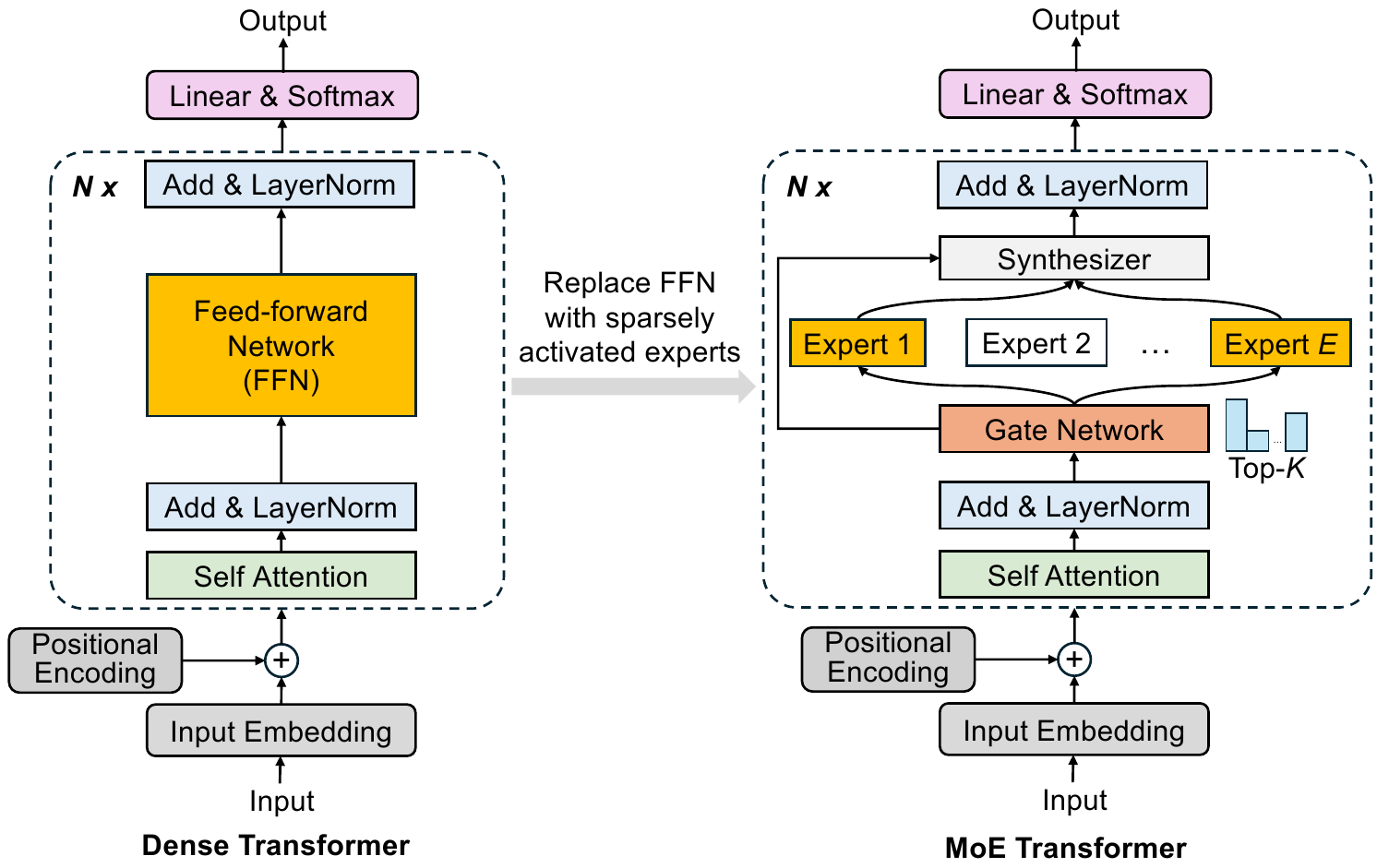}
\caption{Comparison between a dense transformer and an MoE transformer under a typical Top-K routing strategy. The yellow modules denote the activated feed-forward components, including the dense FFN in the dense transformer and the selected experts in the MoE transformer.
\label{fig:Dense_vs_MoE}}
\end{figure}

Fig. \ref{fig:Dense_vs_MoE} compares a dense transformer with an MoE transformer. Both architectures map input tokens into embeddings, combine them with positional encoding, and process them through stacked transformer layers followed by a linear projection and softmax. Their key difference lies in the feed-forward stage of each layer. In a dense transformer, the feed-forward network (FFN) is fully activated for every token. In contrast, in an MoE transformer, the FFN is replaced by multiple experts and a gating network, which computes expert relevance scores and dynamically selects a small subset of experts for each token. A commonly used strategy is Top-K routing, where the K experts with the highest scores are activated. The selected experts process the token representations in parallel, and their outputs are aggregated using the gating weights before being passed to the next layer. In this way, the MoE transformer retains the overall transformer architecture while replacing dense computation with sparse expert activation, thereby reducing per-token computation and improving scalability.
This sparsity also makes MoE particularly suitable for distributed and resource-constrained environments, such as satellite networks.

\subsection{From Traditional Space Inference to SpaceMoE}

For LLM inference, the space system mainly consists of four components: \textit{satellites}, which integrate sensing, communication, and computing capabilities to process and relay data; \textit{ground cloud data centers}, which handle the heavy model pre-training or fine-tuning and then dispatch the LLMs to satellites; \textit{ground stations}, which serve as relay nodes connecting satellites to cloud data centers or terrestrial users; and \textit{terrestrial users}, who utilize or benefit from the satellites' computing results.

As shown in Fig. \ref{fig:centralized_vs_decentralized}, we first elaborate on the traditional inference approaches in satellite networks, and then introduce the SpaceMoE architecture, an expert-level paradigm designed to effectively harness fragmented resources in space.

\subsubsection{Traditional Space Inference}
Space inference can be realized in either centralized or distributed forms, as detailed below.

\paragraph{Ground-based Centralized Inference}
This represents the conventional satellite data processing paradigm, where all tasks are offloaded to ground stations and then forwarded to cloud data centers for centralized MoE inference. Such architecture can leverage abundant terrestrial computing resources without deploying LLMs on satellites. However, it relies heavily on backhaul transmission, which may introduce long latency and limit service availability under intermittent ground connectivity.
 
\paragraph{On-orbit Centralized Inference}
Unlike ground-based centralized inference, on-orbit centralized MoE inference deploys and executes all model components on a single satellite. This avoids inter-satellite communication and keeps both inputs and outputs local. However, all experts must be stored and executed onboard, requiring the full model to reside on one satellite.
This architecture is feasible only for MoE models of small or moderate size, since onboard GPU memory, storage, computing capability, and energy are limited.

\paragraph{Split (Distributed) Inference}
Split inference partitions the model by transformer layers and assigns different layer groups to multiple satellites~\cite{zhang2026communication}. 
While split inference has been widely studied, it suffers from two key limitations: excessive communication overhead and long end-to-end latency. Considering U-shaped split inference, each token’s hidden state must be sent to helper satellites and then returned, leading to a per-token communication cost proportional to twice the hidden-state size. The communication cost becomes even more severe when the model is partitioned as a chain over multiple satellites, or when handling long-context tasks. In addition, since execution is subject to layer-wise dependencies without parallelism, the end-to-end latency can be excessive.

\subsubsection{Expert-level Space Inference (SpaceMoE)}
\begin{figure*}[t]
	\centering
    \includegraphics[width = 1\textwidth]{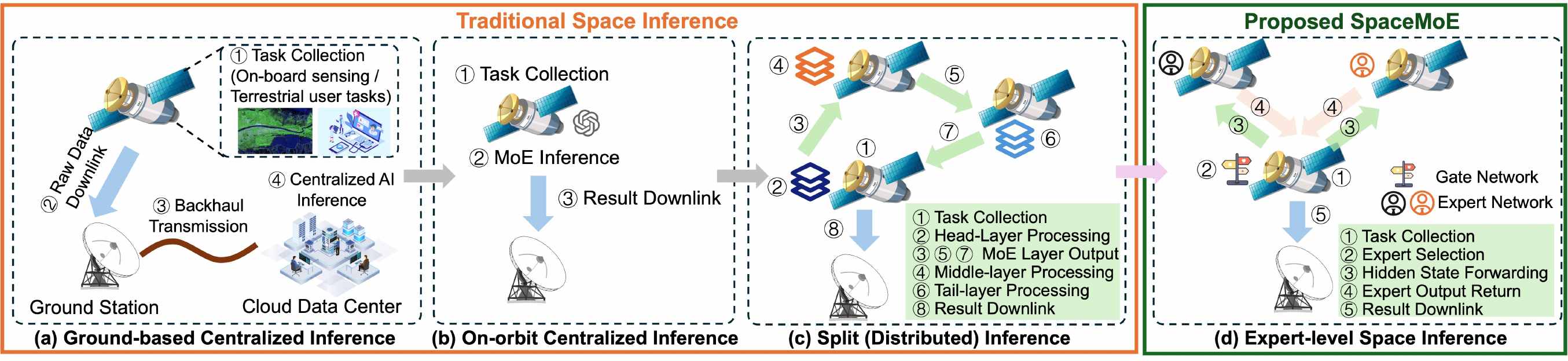}
\caption{Evolution of space general AI inference architectures: From traditional space inference to SpaceMoE.
\label{fig:centralized_vs_decentralized}}
\end{figure*}

To overcome the inflexibility of traditional space inference paradigms, we propose SpaceMoE, an \textit{expert-level} split inference architecture. The key idea is to distribute the expert networks across multiple satellites for parallel computation (see Fig. \ref{fig:centralized_vs_decentralized}). 
This architecture has two salient advantages. First, it is perfectly aligned with the inherent sparsity of MoE models. Since expert activation is typically highly skewed -- especially in deeper layers -- the source satellite can cache frequently used experts locally to drastically reduce inter-satellite communications. Second, it enables parallel computation within each layer on multiple satellites to accelerate inference, in stark contrast to the sequential computation of split inference.

\textbf{Remark}. 
Due to the fine-grained nature of SpaceMoE’s expert-level design, the three traditional space inference approaches discussed above can be viewed as special cases under different deployment configurations. 
Specifically, SpaceMoE naturally reduces into split inference when one or more entire transformer layers are offloaded to a helper satellite, or into centralized inference when local execution is preferred due to high inter-satellite communication latency. As a result, SpaceMoE offers superior flexibility and scalability for distributed MoE inference in satellite networks.

\subsubsection{Comparative Discussion}

Table \ref{tab:comparison_architecture} summarizes the main differences among the three on-orbit inference architectures in terms of computation placement, communication costs, parallelism, memory requirement, and scalability. On-orbit centralized inference is simple but heavily constrained by onboard resources. Split inference supports larger models, but still suffers from substantial communication overhead and limited parallelism. By contrast, SpaceMoE better exploits the sparse structure of MoE models, making it a more scalable and communication-efficient design for satellite networks.

\begin{table}[t]
\caption{Comparison of SpaceMoE and other on-orbit inference architectures}\label{tab:comparison_architecture}
\centering
\begin{tabular}{|C{1.6cm}|C{1.6cm}|C{1.8cm}|C{1.8cm}|}
\hline
\textbf{Aspect} & \textbf{On-orbit Centralized Inference} & \textbf{Split (Distributed) Inference} & \textbf{Proposed SpaceMoE} \\
\hline \hline
Computation placement & Fully local & Layer-wise distributed & Expert-wise distributed \\ \hline
Communication cost & None & Per-token cost scales with helper satellite count & Sparse, on-demand \\ \hline
Parallelism & Low & Low & High \\ \hline
Memory requirement at source satellite & High & Moderate & Low to moderate \\ \hline
Scalability & Low & Moderate & High \\ \hline
\end{tabular}
\end{table}

\section{Key Problems and Solutions in SpaceMoE}\label{sec:prob_solution}

Building on the expert-level split MoE inference architecture introduced in Section \ref{sec:architecture}, we next discuss three key research problems in SpaceMoE, namely expert placement, expert selection, and hidden-state transmission and routing schemes.

\subsection{Expert Placement in SpaceMoE}
\subsubsection{Problem Statement}
To enable SpaceMoE, the very first problem is how to determine expert placement, i.e., how experts are placed and replicated across satellites under limited onboard memory/storage. This is long-term system design, since it determines where experts reside in the satellite network and directly affects subsequent expert access during distributed inference. In general, the goal of expert placement is to improve expert accessibility and reduce the communication cost of remote expert invocation.

\subsubsection{Challenges}
Compared with terrestrial distributed MoE systems, expert placement in SpaceMoE is fundamentally reshaped by the dynamic nature of satellite networks. In terrestrial systems, expert placement is typically optimized to reduce inference latency and improve accessibility for geographically distributed users under relatively stable network conditions~\cite{11395617}. By contrast, in satellite networks, expert accessibility is inherently time-varying because satellite mobility continuously changes network topology and inter-satellite connectivity. As a result, an expert that is easy to access at one moment may become difficult to reach later due to disrupted links or longer multi-hop paths. This makes static placement strategies developed for terrestrial systems insufficient for SpaceMoE.

Constellation scale also plays an important role in expert placement. A small constellation limits the available satellites for hosting and replicating experts, which may reduce expert accessibility during certain orbital periods. A large constellation offers more placement and replication choices, but also increases the search space and coordination overhead under the time-varying topology. Therefore, expert placement in SpaceMoE should be scale-aware, balancing improved accessibility with placement complexity.

Another challenge comes from the activation structure of MoE models. In practice, expert activation is often highly skewed, meaning that a small subset of experts is selected much more frequently than others. In addition, multiple experts may be co-activated during inference. In space environments, these properties interact strongly with network dynamics: if frequently activated experts are placed on poorly connected satellites, repeated remote access can cause substantial communication overhead; similarly, if frequently co-activated experts are placed far apart, their joint execution may incur significant latency. Therefore, expert placement in SpaceMoE must jointly account for both time-varying connectivity and the activation/co-activation structure of the model.

\subsubsection{Solutions}
\begin{figure}[t]
	\centering
    \includegraphics[width = 0.4\textwidth]{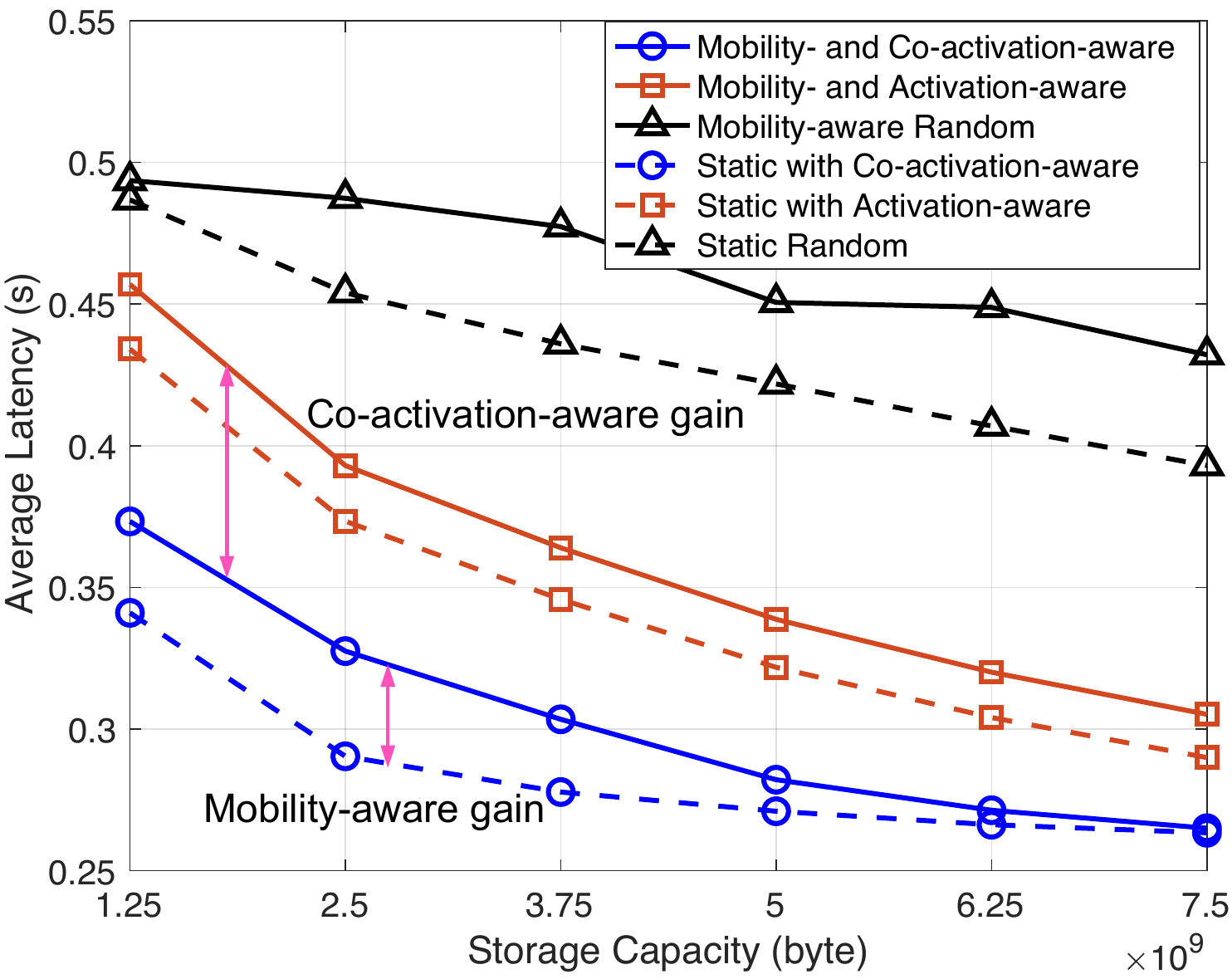}
    \caption{Average latency per token in satellite networks, where the ground user sends its MoE inference request to satellites for on-orbit processing. The satellite-related parameters are set according to the Starlink constellation, while the model and computation parameters follow \cite{11395617}.
\label{fig:space_expert_placement}}
\end{figure}

These observations motivate mobility-aware expert placement strategies for SpaceMoE. First, expert replicas should be placed by considering both the temporal evolution of satellite connectivity and the constellation scale, so that frequently requested experts remain accessible over time rather than only under a fixed snapshot of the network. Specifically, small constellations require more selective expert placement due to limited hosting opportunities, while large constellations allow more flexible replication by exploiting richer satellite candidates.
Second, due to the skewed activation distribution, it is more effective to selectively replicate a small set of highly activated experts instead of uniformly replicating all experts. Third, experts that are often co-activated should be co-located or placed nearby whenever possible, so that joint inference can be completed with lower inter-satellite communication cost. Overall, expert placement in SpaceMoE should move beyond static placement and instead jointly exploit satellite mobility patterns, constellation scale, and MoE activation statistics to support efficient distributed inference.
Fig. \ref{fig:space_expert_placement} shows the average latency per token under different expert placement strategies in satellite networks.
It indicates that mobility-aware placement outperforms static placement and incorporating co-activation information brings further latency reduction. This highlights the need to jointly exploit satellite mobility and MoE activation statistics in expert placement.

\subsection{Expert Selection in SpaceMoE}
\subsubsection{Problem Statement}
Given expert placement, it is also crucial to optimize expert selection in SpaceMoE, i.e., determining which experts are activated for each token during inference. This is a short-term decision that directly affects both inference performance and system cost, since it not only determines model execution paths but also implicitly decides which expert-hosting satellites participate in inference.

\subsubsection{Challenges}
Compared with terrestrial distributed MoE systems, expert selection in SpaceMoE is fundamentally reshaped by the solar-powered and battery-supported energy system of satellites. In terrestrial systems, expert selection is usually optimized according to gating scores, channel conditions, or expert similarity, with the main objective of balancing inference accuracy and communication efficiency~\cite{chen2026siftmoe}. By contrast, in satellite networks, activating an expert not only incurs communication-computing costs but also triggers significant energy consumption. When harvested solar power is insufficient, this computation requires battery discharge, which induces irreversible battery degradation and affects satellite lifetime.

Moreover, the resulting degradation is not determined only by total energy consumption. It also depends on the depth and rate of discharge, making the degradation process inherently nonlinear and state-dependent~\cite{zeng2026unseen}. As a result, the same expert computation may incur very different long-term costs under different operating conditions, such as sunlight or eclipse periods. Therefore, unlike terrestrial expert selection, which mainly focuses on immediate inference utility and communication cost, expert selection in SpaceMoE must additionally account for the degradation-aware execution cost induced by expert activation.

\begin{figure}[!t]
	\centering
	\subfigure[Model: Switch-base-8. Dataset: XSum.]{\includegraphics[width =0.24\textwidth]{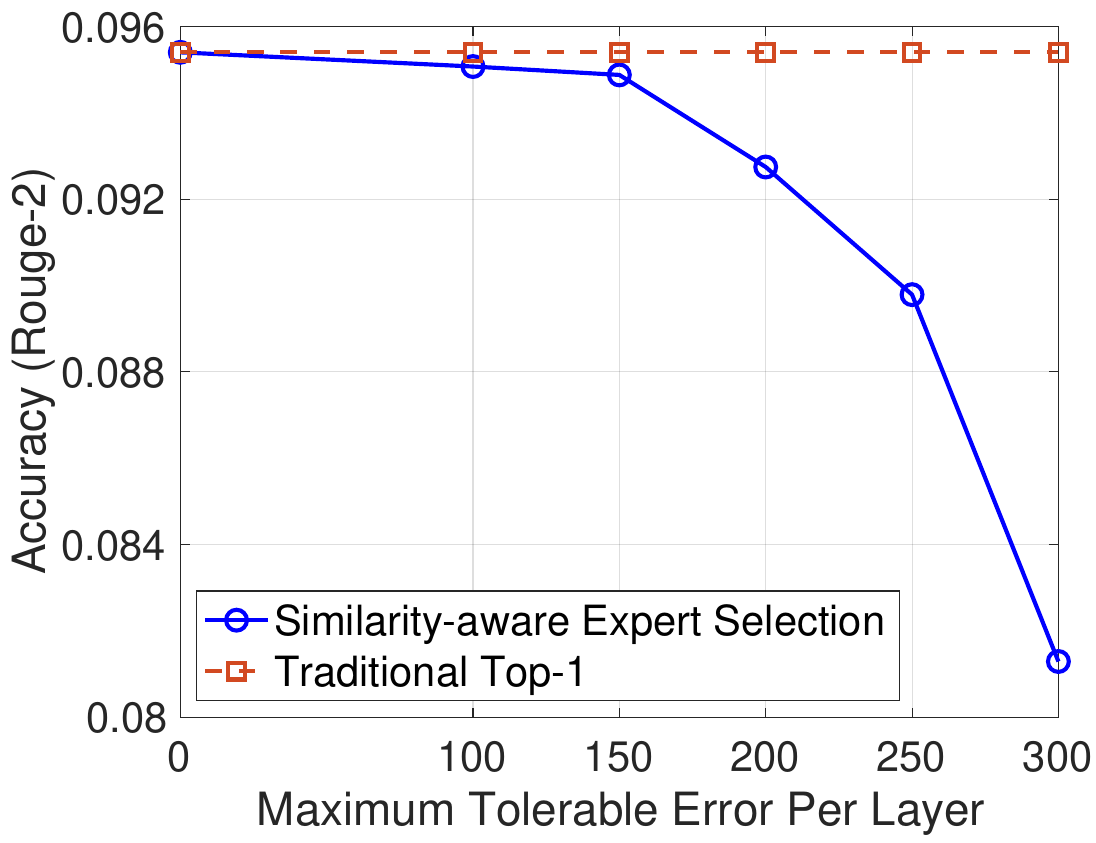}\label{fig:ExpertSelection_top1}}
	\subfigure[Model: Mixtral-8x7B. Dataset: CommonsenseQA.]{\includegraphics[width =0.24\textwidth]{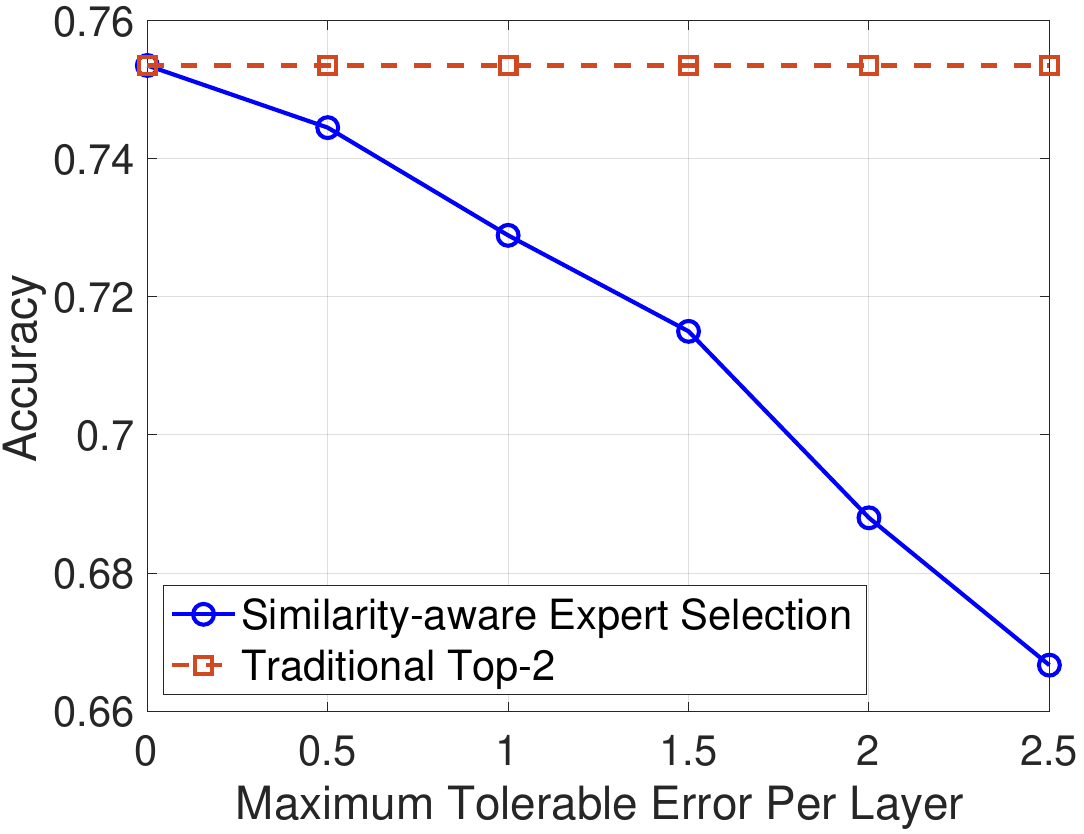}\label{fig:ExpertSelection_top2}}
	\caption{Accuracy comparison between similarity-aware expert selection and traditional Top-$K$ routing under different error tolerance levels.} 
\label{fig:expert_selection} 
\end{figure}

\subsubsection{Solutions}
Fig. \ref{fig:expert_selection} provides an important insight for SpaceMoE expert selection. The results show that similarity-aware expert selection can keep the accuracy degradation within a limited range. 
These observations motivate a battery-degradation-aware expert selection strategy for SpaceMoE. Rather than selecting experts solely according to model importance or communication efficiency, the system should also account for the marginal battery degradation caused by the corresponding computation. In particular, when multiple candidate experts provide similar utility, the system can favor those hosted on satellites with more favorable energy conditions or healthier battery states, thereby reducing long-term degradation. Overall, expert selection in SpaceMoE should balance short-term inference performance with long-term system sustainability.

\subsection{Hidden-State Transmission and Routing in SpaceMoE}

\subsubsection{Problem Statement}

Given expert placement and expert selection, the next problem is how to transmit hidden states between the source satellite and the selected expert-hosting satellites (i.e., expert satellite). In distributed SpaceMoE inference, once an expert is activated for a token, the corresponding hidden state must be routed from the source satellite to the expert satellite, and the expert output must be returned for subsequent inference. Since multiple inter-satellite paths may exist between the source satellite and the expert satellite, hidden-state transmission and routing directly affect communication costs and inference reliability. Therefore, the goal is to design routing and transmission strategies that can efficiently deliver hidden states under dynamic satellite connectivity, lossy space links, and satellite resource constraints.

\subsubsection{Challenges}

Compared with terrestrial distributed MoE systems, hidden-state transmission in SpaceMoE is fundamentally affected by the orbital-layer structure of satellite networks. In a single-layer LEO constellation, hidden states are mainly transmitted through intra-layer ISLs with relatively short propagation delay but limited and time-varying path choices. In a multi-layer constellation, higher-orbit satellites can provide wider coverage and more stable relay opportunities, but at the cost of longer propagation delay and cross-layer transmission overhead. Therefore, hidden-state routing should account for the constellation-layer structure rather than relying only on conventional latency- or bandwidth-oriented routing.

Another challenge comes from the importance heterogeneity of hidden-state transmissions. Hidden states associated with different experts may have different impacts on inference accuracy, and thus require different reliability and compression levels. Such heterogeneity should be jointly considered with the constellation-layer structure: critical hidden states may be routed through more stable cross-layer paths, while less critical ones can use lower-cost intra-layer paths or higher compression.
Therefore, hidden-state transmission should jointly consider path conditions, expert importance, and constellation-layer structure rather than treating all transmissions equally.

\begin{figure}[t]
	\centering
    \includegraphics[width = 0.4\textwidth]{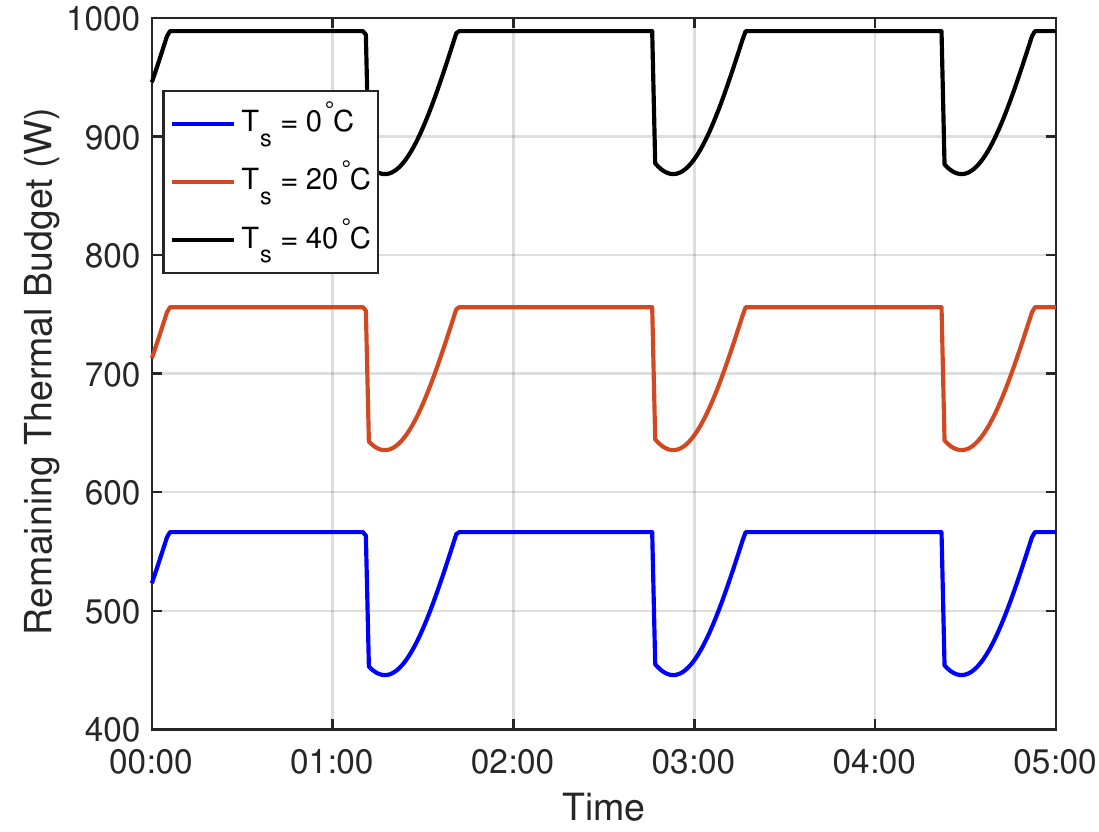}
\caption{Remaining thermal budget versus time under different radiator temperatures $T_s$. The parameters are set according to \cite{starcloud2024spaceai}.
\label{fig:thermal_budget}}
\end{figure}

Furthermore, routing decisions are constrained by the time-varying thermal budget of satellites along the path. Unlike terrestrial platforms with conventional cooling mechanisms such as convection or liquid cooling, satellites primarily dissipate heat through radiation, which limits onboard communication and computation. Denoting the absorbed environmental heat and radiated thermal power by $P_{\rm abs}$ and $P_{\rm rad}$, respectively, the remaining thermal budget can be written as $P_{\rm budget}=P_{\rm rad}-P_{\rm abs}$. This budget changes with orbital motion, sunlight–eclipse transitions, and Earth-induced radiation, as shown in Fig. \ref{fig:thermal_budget}. Therefore, even when a satellite has sufficient energy, it may not have enough thermal margin to support additional forwarding or expert execution.

\subsubsection{Solutions}

These observations motivate hidden-state transmission and routing strategies in SpaceMoE that jointly account for the constellation-layer structure, expert importance, and satellite thermal conditions. First, routing decisions should account for the orbital-layer structure of satellite networks. Specifically, intra-layer paths can be preferred for latency-sensitive hidden-state transmissions, while cross-layer paths can be exploited when more stable relay opportunities are needed. Second, hidden-state transmission should be importance-aware. For hidden states associated with important experts, the system can select more reliable paths or use lower-compression transmission to preserve inference accuracy. For less important expert activations, the system can choose lower-cost paths or apply more aggressive compression to reduce communication overhead.
Third, routing should also account for the thermal conditions of relay and expert-hosting satellites. Instead of always choosing the shortest path, the system can avoid satellites with limited remaining thermal budgets and prefer paths whose relay and destination satellites have sufficient thermal margins. 

\subsection{Design Tradeoffs in SpaceMoE}
    Taken together, the above discussions show that SpaceMoE is shaped by several distinctive design tradeoffs that do not arise in the same form in terrestrial systems. In particular, the key challenges of expert placement, selection, transmission, and routing can be summarized through the following tradeoffs.
    
\begin{itemize}
    \item \textbf{Access latency vs. storage efficiency.} Expert placement must cope with time-varying topology and changing inter-satellite connectivity. Replicating experts across more satellites can improve accessibility and robustness, but it also consumes limited onboard storage/memory. Therefore, expert placement must balance connectivity robustness with storage efficiency, motivating selective replication and co-location of frequently used experts.

    \item \textbf{Short-term inference performance vs. long-term battery sustainability.}     Activating an expert not only affects inference utility and communication cost, but also triggers onboard computation on the hosting satellite. When harvested solar power is insufficient, such computation requires battery discharge, which induces nonlinear and state-dependent degradation. Therefore, expert selection must balance immediate inference performance with long-term battery health.
    
    \item \textbf{Inference accuracy vs. thermal efficiency.} More reliable transmission paths can reduce hidden state distortion and improve inference accuracy, especially for critical expert activations. However, they may also consume more communication resources. Since satellites have limited and time-varying thermal budgets, routing through thermally constrained satellites may make hidden-state forwarding or expert execution infeasible. Therefore, hidden-state transmission and routing in SpaceMoE must balance inference accuracy with communication and thermal efficiency under dynamic orbital conditions.
\end{itemize}

\section{Open Problems in SpaceMoE}\label{sec:openprob}

Despite the significance of SpaceMoE, its practical realization in satellite networks is still at an early stage. Beyond the core design problems of expert placement, expert selection, and hidden-state transmission and routing, several broader issues remain open and deserve further investigation to fully unlock the potential of SpaceMoE systems.

\begin{itemize}
    \item     \textbf{Security and trustworthiness in SpaceMoE.} In SpaceMoE, hidden states, expert outputs, and model parameters may be exchanged across multiple satellites during distributed inference, creating a much larger attack surface than centralized execution. This raises security risks such as eavesdropping, model stealing, malicious expert manipulation, and inference corruption. Moreover, the distributed and harsh space environment makes conventional terrestrial trust assumptions less applicable. This calls for SpaceMoE-specific mechanisms, including secure expert invocation, trusted inter-satellite collaboration, privacy-preserving hidden-state exchange, and robust inference under compromised nodes.

    \item \textbf{Expert Quantization in SpaceMoE.} Expert quantization determines the numerical precision used to store and execute different experts across satellites. Higher-precision expert execution usually incurs heavier memory access and computation load, thereby generating more heat on the hosting satellite. Since the remaining thermal budget of satellites varies with orbital conditions, fixed quantization strategies may be insufficient for SpaceMoE. This calls for SpaceMoE-specific quantization mechanisms that jointly consider expert importance, inference accuracy, onboard resource constraints, and dynamic thermal conditions.

\item \textbf{On-orbit fine-tuning of SpaceMoE models.} SpaceMoE models are typically trained on the ground data center and then deployed to satellites for on-orbit MoE inference. However, the dynamic service environment, changing task distributions, and region-specific user demands may require periodic fine-tuning. A straightforward solution is to exchange model updates through ground stations or ISLs, but this may introduce substantial communication overhead. This calls for communication-efficient on-orbit fine-tuning mechanisms for SpaceMoE. One promising direction is to exploit the mobility and store-carry-forward capability of satellites~\cite{11314800}. The key idea is to enable cross-satellite and cross-region parameter mixing only when inter-satellite/user-satellite connectivity is strong, thus facilitating collaborative learning in space with minimum communication and bandwidth costs.

\end{itemize}

	\section{Conclusion Remarks}\label{sec:conclusion}
	In this article, we have presented SpaceMoE as a promising paradigm for enabling scalable distributed LLM inference in satellite networks. Different from terrestrial distributed MoE systems, SpaceMoE is fundamentally shaped by satellite-specific constraints, including dynamic topology, battery degradation, and time-varying thermal conditions. To provide a structured understanding of this emerging direction, we reviewed its industrial and standardization motivations, introduced the architectural evolution of SpaceMoE, and discussed three core design problems in SpaceMoE, namely expert placement, expert selection, and hidden-state transmission and routing. We further highlighted several important open problems that may play a key role in its future development. Overall, SpaceMoE offers a new path toward realizing efficient and sustainable on-orbit general intelligence, and we hope this article will stimulate further research on distributed LLM inference in future satellite networks.
    
\ifCLASSOPTIONcaptionsoff
	\newpage
	\fi

	\bibliographystyle{IEEEtran}
	\bibliography{IEEEabrv,body/reference.bib}

    \section*{Biographies}
   
    \vspace{-2.5cm}
    \begin{IEEEbiographynophoto}{Qian Chen} is a postdoctoral fellow at the Department of Electrical and Computer Engineering, The University of Hong Kong. Her research interests include wireless networking, edge AI, and space-air-ground integrated networks.
\end{IEEEbiographynophoto}

\vspace{-2.5cm}
\begin{IEEEbiographynophoto}{Xianhao Chen} is an assistant professor at the Department of Electrical and Computer Engineering, The University of Hong Kong. His research interests include wireless networking, edge intelligence, and distributed learning.
\end{IEEEbiographynophoto}

\vspace{-2.5cm}
\begin{IEEEbiographynophoto}{Min Sheng} [Fellow, IEEE] is a Full Professor and the Director with the State Key Laboratory of Integrated Service Networks, Xidian University. Her research interests include mobile ad hoc networks, 5G mobile communication systems, and satellite communications networks. 
\end{IEEEbiographynophoto}

\vspace{-2.5cm}
\begin{IEEEbiographynophoto}{Kaibin Huang} [Fellow, IEEE] is a professor at the Department of Electrical and Computer Engineering, The University of Hong Kong. His research interests include mobile edge computing, edge AI, and 6G systems.
\end{IEEEbiographynophoto}
    
\end{document}